\documentclass[lettersize,journal]{IEEEtran}

\usepackage[T1]{fontenc}
\usepackage[utf8]{inputenc}
\usepackage{amsmath,amsthm,mathtools,amssymb}
\usepackage{graphicx}
\usepackage{xcolor}
\usepackage{todonotes}
\usepackage{siunitx}

\usepackage{url}

\usepackage[breaklinks]{hyperref}
\usepackage{orcidlink}

\begin{document}

\title{Impact of Market Reforms on Deterministic Frequency Deviations in the European Power Grid}

\author{%
Philipp C.~Böttcher\orcidlink{0000-0002-3240-0442}, %
Carsten Hartmann\orcidlink{0009-0007-5067-589X}, %
Andrea Benigni\orcidlink{0000-0002-2475-7003}, \textit{Senior Member, IEEE}, %
Thiemo Pesch\orcidlink{0000-0002-3297-6599},
Dirk Witthaut\orcidlink{0000-0002-3623-5341}
\thanks{
The authors are with the Forschungszentrum J\"ulich, Institute of Climate and Energy Systems: Energy Systems Engineering (ICE-1),  52428 J\"ulich, Germany. 
A.~Benigni is with the RWTH Aachen University, Aachen, 52062, Germany. D.~Witthaut is with the
Institute for Theoretical Physics, University of Cologne, 50937 K\"oln, Germany.
}
}

\maketitle

\begin{abstract}
Deterministic frequency deviations (DFDs) are systematic and predictable excursions of grid frequency that arise from synchronized generation ramps induced by electricity market scheduling. In this paper, we analyze the impact of the European day-ahead market reform of 1 October 2025, which replaced hourly trading blocks with quarter-hourly blocks, on DFDs in the Central European synchronous area. Using publicly available frequency measurements, we compare periods before and after the reform based on daily frequency profiles, indicators characterizing frequency deviations, principal component analysis, Fourier-based functional data analysis, and power spectral density analysis. We show that the reform substantially reduces characteristic hourly frequency deviations and suppresses dominant spectral components at hourly and half-hourly time scales, while quarter-hourly structures gain relative importance. 
While the likelihood of large frequency deviations decreases overall, reductions for extreme events are less clear and depend on the metric used.
Our results demonstrate that market design reforms can effectively mitigate systematic frequency deviations, but also highlight that complementary technical and regulatory measures are required to further reduce large frequency excursions in low-inertia power systems.
\end{abstract}

\section{Introduction}

The ongoing energy transition poses fundamental challenges to the operation and stability of electric power systems. An increasing share of renewable generation is connected to the grid via power electronic inverters, which provide little or no intrinsic rotational inertia. As a result, the aggregate system inertia is declining, and frequency stability is becoming increasingly at risk~\cite{milano2018foundations}.

\begin{figure}[tb]
    \centering
    \includegraphics[width=\columnwidth]{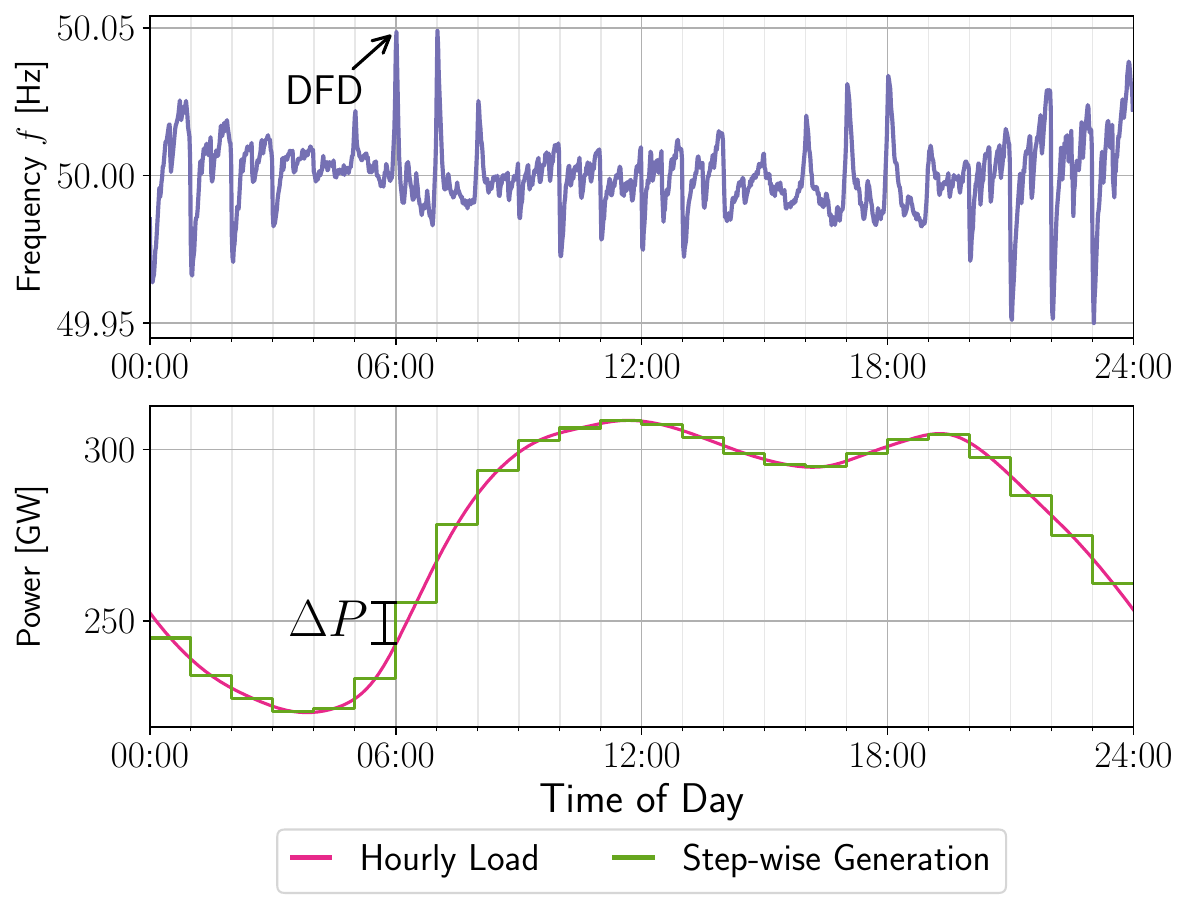}
    \caption{
    Emergence of hourly deterministic frequency deviations (DFDs) due to electricity trading.
    Upper panel: The daily profile of the grid frequency in the Continental European synchronous area shows characteristic peaks at the beginning of every hour.
    Lower panel: These DFDs can be understood from the design of European electricity markets. Load evolves continuously, while generation is traded in blocks. At the beginning of each block, generation is adjusted in an approximately stepwise manner, leading to a power imbalance $\Delta P$ and thus to DFDs. Until 1 October 2025, electricity was mainly traded in hourly blocks, resulting in pronounced hourly DFDs.
    The figure shows the daily profile of the load as a continuous curve, while generation is depicted as a step function for illustrative purposes. The illustration uses data from 2019.
    }
    \label{fig:dfd-market}
\end{figure}

Deterministic frequency deviations (DFDs) are predictable and repeatable departures of the system frequency from its nominal value, caused by systematic imbalances between power generation and load. 
The mechanism underlying DFDs in the Central European power system was summarized in Ref.~\cite{weissbach2009high} and is illustrated in Fig.~\ref{fig:dfd-market}. Electricity trading is organized in hourly or quarter-hourly blocks, which induces strong generation ramps at the boundaries of the trading intervals.
The resulting temporary overgeneration (or scarcity) leads to a rapid increase (or decrease) in system frequency before the imbalance is compensated by the load–frequency control system.

DFDs have been identified by ENTSO-E, the official representative of the European transmission system operators, as a major challenge to system operation, leading to the publication of a comprehensive assessment report in 2020~\cite{entsoe_dfd2020}. The report highlights that DFDs can systematically activate frequency containment reserves, induce unscheduled power flows, and reduce effective damping margins, thereby weakening system operation even in the absence of major contingencies. These effects make DFDs a relevant operational concern and motivate the need for targeted mitigation measures.

The design of the European electricity market plays a central role in the emergence of DFDs. Traditionally, the day-ahead market was organized in hourly trading blocks, leading to pronounced generation ramps at the beginning of each hour, particularly during morning and evening periods (see Fig.~\ref{fig:dfd-market}). 
Intraday electricity trading in Germany was introduced in the early 2010s and further developed with the introduction of quarter-hourly intraday auctions in 2014, enabling smaller, more frequent generation adjustments. Indeed, modest deviations at half-hourly and quarter-hourly intervals can already be observed in the daily frequency profile shown in Fig.~\ref{fig:dfd-market}. However, due to its significantly lower trading volume~\cite{han2022complexity}, the intraday market could not fully mitigate the hourly DFDs.
The day-ahead market was revised in 2025, replacing hourly blocks with quarter-hourly blocks and thereby directly addressing one of the structural causes of DFDs~\cite{eu2025electricity}. The market reform took effect on 30 September for delivery day 1 October 2025. 

In this article, we investigate the impact of this market design reform on DFDs in the Central European power system. We compare frequency time series before and after the reform using stability indicators as well as hourly and quarter-hourly patterns. In this way, we quantify how the market reform has reduced deterministic frequency deviations.

This article is organized as follows. We review the scientific literature on DFDs in Sec.~\ref{sec:literature}. The methods used for the analysis of frequency time series are described in Sec.~\ref{sec:methods}. The results are presented in Sec.~\ref{sec:results} and discussed in Sec.~\ref{sec:discussion}.

\section{Literature Review}
\label{sec:literature}

The origins of market-induced DFDs in the Continental European (CE) grid, as sketched in Fig.~\ref{fig:dfd-market}, were summarized by Weissbach and Welfonder in Ref.~\cite{weissbach2009high}. The authors identify block-based market scheduling as a key driver of synchronized power ramps and propose the introduction of quarter-hourly trading intervals as a suitable countermeasure. A more detailed analysis of the impact of generator ramps on DFDs was given in ~\cite{weissbach2008improvement}. Borsche et al.~extended this analysis to control signals and proposed a new framework for control reserves based on batteries and related units~\cite{borsche2014new}, while the fundamental explanation of DFDs was refined in Ref.~\cite{kruse2021exploring}. Solar power, due to its comparatively slow ramping behavior, can mitigate DFDs and should therefore be excluded from the power ramps illustrated in Fig.~\ref{fig:dfd-market}.

The ENTSO-E Report on Deterministic Frequency Deviations provides a comprehensive empirical and system-level analysis of DFDs in the CE synchronous area, combining long-term frequency measurements, statistical trend analyses, and detailed investigations of root causes and mitigation options~\cite{entsoe_dfd2020}. The report systematically attributes DFDs primarily to structural features of market scheduling and ramping behavior, quantifies their impact on reserves, power flows, and system damping, and evaluates a wide range of technical and market-based countermeasures through simulations and national case studies. Beyond documenting the phenomenon, the report establishes operational performance targets, monitoring approaches, and a policy-relevant framework for assessing trade-offs between market design reforms and reserve-based mitigation strategies.

Market design and institutional factors play a crucial role in shaping balancing needs and frequency control efforts, sometimes outweighing the impact of increasing variable renewable generation~\cite{hirth2013balancing}. A central mechanism underlying market-induced DFDs is the use of block-based market time units, which can trigger synchronized schedule changes and amplified system-wide ramping~\cite{meeus2020evolution}. Coarse temporal resolution in market schedules therefore increases redispatch needs and operational stress, motivating reforms toward finer market time units~\cite{neuhoff2015power}.  
In principle, reducing market time units weakens this synchronization and mitigates deterministic imbalances. Empirical evidence from the intraday market supports this view: the introduction of quarter-hourly intraday products and tighter balance-responsibility rules in Germany substantially reduced deterministic imbalances and frequency restoration reserve demand, although residual effects persist~\cite{weissbach2018impact}. Using frequency time-series data, Schäfer et al. further demonstrate that shorter trading intervals reduce the likelihood of large frequency deviations compared to coarser, discrete scheduling~\cite{schafer2018isolating}.

From a balancing-market perspective, Motte-Cortés and Eising~\cite{motte2019assessment} document a Europe-wide shift towards finer temporal resolution and closer-to-real-time operation, driven by increasing renewable penetration. Looking ahead, ter Borg et al.~argue that further reducing market time units to five minutes could markedly decrease systematic imbalances in future high-renewable systems, albeit at the cost of increased complexity for market participants~\cite{ter2024should}. Together, these studies indicate that while finer market time units are a promising lever to mitigate DFDs, their interaction with system inertia, control reserves, and frequency dynamics remains an open research challenge.

General aspects of frequency dynamics and control in future low-inertia grids have been reviewed in Ref.~\cite{milano2018foundations}. A model-based analysis of the impact of step changes in power generation was presented in Ref.~\cite{ulbig2014impact}. The impact of increasing inverter-based generation on frequency nadir and RoCoF has also been studied using large-scale dynamic simulations~\cite{gevorgian2014investigating}.
A data-driven analysis of frequency deviation indicators based on explainable machine learning confirms the importance of rapid generation ramps, in particular hydropower~\cite{kruse2021revealing}. A physics-informed machine learning model predicting both DFDs and ambient frequency fluctuations was presented in Ref.~\cite{kruse2023physics}. Notably, other recurrent grid-frequency patterns exist at shorter periods that are not driven by market mechanisms~\cite{mishra2025understanding,hartmann2025cyber}.

\section{Methods}
\label{sec:methods}

We analyze frequency time series $f(t)$ from the Central European power system before and after the market reform taking effect on 1 October 2025. We compare the power spectral density, the daily frequency profile and indicators characterizing frequency deviations, such as the rate of change of frequency (RoCoF), as described below. Finally, we quantify the contribution of quarter-hourly and hourly patterns in the time series via principal component analysis (PCA) and Fourier-based functional data analysis (FDA). 

\subsection{Data source}
\label{sec:data_source}

Frequency data are publicly provided by the German transmission system operators~\cite{Netztransparenz_data} at a temporal resolution of $\tau = 1\,\si{s}$. All data is reported in German local time. For the analysis, we center the frequency around the reference value, $\tilde f(t) = f(t) - f_{\rm ref}$ with $f_{\rm ref} = 50\,\si{Hz}$.

Throughout the paper, we compare data for three months before and after the market reform. Specifically, we consider the period from 1 October 2025 at 00:00 to 31 December 2025 at 00:00 to the respective period in 2024, to exclude seasonal effects as much as possible. 
For selected analyses, additional data from earlier years are included to place the observed changes in a longer-term context (cf.~Fig.~\ref{fig:long-term}).

We note that data handling and publication procedures at the Netztransparenz platform~\cite{Netztransparenz_data} changed in June 2022, as indicated by the grey shading in Fig.~\ref{fig:long-term}. Frequency data prior to 2022 frequently exhibit data-quality issues, including missing values and inconsistencies. To mitigate these effects, we apply a postprocessing procedure following~\cite{kruse2020predictability} and remove values that either have an unrealistically large deviation from the reference frequency or an unrealistically large difference to previous values. Despite this correction, the data for 2019 remain incomplete, with more than 10\% missing values after postprocessing, and are therefore excluded from the analysis.

We remark that regional differences in the grid frequency are typically small and damped out in seconds~\cite{rydin2020open,rydin2022phase} such that the results derived from this data are representative for the entire synchronous area.

\subsection{Power spectral density}
\label{sec:method_psd}

To analyze the spectral properties of the grid frequency $f(t)$, we compute the power spectral density $S(1/T)$, which quantifies how the power of $f(t)$ is distributed at different frequency-domain components with the period $T$. Comparing $S(1/T)$ before and after the market reform reveals spectral changes across all timescales.

We estimate $S(1/T)$ using Welch’s method~\cite{welch1967use} implemented in SciPy~\cite{2020SciPy-NMeth}. 
The signal is split into overlapping segments, detrended, and a Hann-window is applied to each segment to improve the distinction of neighboring peaks.
Subsequently, each segment is transformed via the discrete Fourier transform, and the spectra of segments are averaged to obtain a good estimate of the power spectral density.

\subsection{The daily profile}
\label{sec:method-daily-profile}

The daily profile of the grid frequency allows for a straightforward visual analysis of recurrent patterns. Splitting time $t$ into the day index $d$ and the time of day $t'$, the daily profile is defined as
\begin{align}
    f_{\mathrm{dp}}(t') = \frac{1}{N_d} \sum_{d} f(d,t'),
    \label{eq:def-daily}
\end{align}
where the sum is taken over all days in the respective period and $N_d$ denotes the number of available days.

In addition to the mean profile, we analyze the empirical $5\%$ and $95\%$ quantiles at each time of day $t'$. These quantiles are defined as the values below (above) which $5\%$ ($95\%$) of the frequency observations at the same time of day across all days are located, and they characterize the variability and asymmetry of recurrent frequency deviations.

\subsection{Indicators of frequency deviations}
\label{sec:methods-indicators}

We consider four indicators characterizing frequency deviations for every hourly interval, following Ref.~\cite{kruse2021revealing}. The intervals are denoted as $I_h$ starting at time $t_h$. These indicators quantify the magnitude, persistence, and temporal structure of frequency deviations during normal system operation and are thus used in the analysis of recurrent DFDs.

The nadir is defined as the largest frequency deviation from the reference value within the interval $I_h$ and is defined as
\begin{align}
    \mathrm{Nadir}(h) = \max_{t \in I_h} |\tilde f(t)| \, .
\end{align}
The mean square deviation (MSD) quantifies the primary control effort in the respective interval~\cite{tyloo2020primary} and is defined as
\begin{align}
    \mathrm{MSD}(h) = \frac{1}{|I_h|} \sum_{t \in I_h} {\tilde f}^2(t).
\end{align} 
The integrated frequency deviation is defined as
\begin{align}
    \mathrm{Integral}(h) = \tau \sum_{t \in I_h} \tilde f(t)
\end{align} 
and indicates a systematic imbalance between power generation and load. 
Finally, we consider the Rate of Change of Frequency (RoCoF) at the beginning of the respective interval.  We estimate the derivative $\frac{df}{dt}$ at time $t$ by smoothing the increments $\delta f(t') = [f(t') - f(t' - \tau)]/\tau$ over an interval $[t-L,t+L]$. 
We then consider the steepest slope of the derivative within a window $W_h = [t_h-T, t_h+T]$ around $t_h$,
\begin{align*}
    \mathrm{RoCoF}(h) =
         \max_{t \in W_h} \left|  \frac{df}{dt} (t) \right| \, .
\end{align*}
We choose $L=T=60 \, \si{s}$ following Ref.~\cite{kruse2021revealing}.
Notably, we consider only the magnitude of the nadir and the RoCoF in the following, i.e.,~we do not distinguish positive and negative frequency deviations in the statistical analysis.

\subsection{Principal component analysis of hourly frequency profiles}

To identify and characterize recurrent patterns in the grid frequency beyond simple averages, we apply principal component analysis (PCA) to hourly segments of the frequency time series. PCA is a standard dimensionality-reduction technique that represents a set of time-dependent signals as a weighted superposition of orthogonal modes, ordered by the amount of variance they explain~\cite{jolliffe2002principal}.

As before, we divide the frequency time series into hourly segments and denote the frequency trajectory within hour $h$ by $f_h(s)$, where $s=1,\dots,3600$ indexes the seconds within the hour. Furthermore, we remove the mean over all segment for each $s$ to obtain $\hat f_h(s)$. PCA yields a decomposition of the form
\begin{align}
    \hat f_h(s) = \sum_{n} a_{h,n} \, v_n(s),
    \label{eq:pca}
\end{align}
where $v_n(s)$ are the orthonormal principal components (modes) and $a_{h,n}$ are the corresponding coefficients for hour $h$. The components $v_n(s)$ are empirically obtained from the data and represent orthogonal intra-hour frequency patterns, ordered by the fraction of total variance they explain, while the coefficients $a_{h,n}$ quantify how strongly these patterns are expressed in individual hours. 

By comparing the leading components and their coefficients before and after the market design reform, we assess how characteristic market-induced frequency patterns change in structure and importance. We note that each principal component is defined only up to an overall sign. Consequently, a given component can represent both positive and negative frequency excursions, depending on the sign of the corresponding coefficient $a_{h,n}$.

\subsection{Fourier-based data analysis}
\label{sec:methods-fourier}

To quantify the relative importance of intra-hour temporal patterns at specific time scales, we complement the PCA-based analysis with a Fourier-based functional data analysis. As above, the frequency time series is partitioned into hourly segments, each treated as a functional observation. 

Within each segment $h$, the centered frequency signal $\tilde f_h(s)$ is represented by a truncated Fourier expansion,
\begin{align}
\tilde f_h(s)
&= c_{h,0}+
\sum_{k \in \mathcal{K}}
\bigg[
c_{h,k}\cos\!\left(\frac{2\pi k s}{T}\right)
\nonumber \\
& \qquad \qquad \qquad \quad +
b_{h,k}\sin\!\left(\frac{2\pi k s}{T}\right)
\bigg]
+
\varepsilon_h(s),
\label{eq:Fourier-expand}
\end{align}
where $\mathcal{K}$ denotes a predefined set of harmonics, $\varepsilon_h(s)$ captures residual fluctuations, and $c_{h,0}$ is the segment mean. Subtracting the segment mean before the decomposition removes constant offsets and suppresses leakage from very low-frequency components.

In contrast to principal component analysis, which extracts data-driven basis functions, the Fourier representation employs a fixed orthogonal basis, allowing individual harmonics to be directly associated with specific intra-hour time scales. The coefficients $(b_{h,k}, c_{h,k})$ are estimated independently for each interval and quantify the amplitude of oscillatory patterns at harmonic $k$.

For each harmonic $k$, we define the corresponding oscillatory energy as
\begin{align}
   E_{h,k} = b_{h,k}^2 + c_{h,k}^2 .
\end{align}
Variability on the hourly scale corresponds to the fundamental mode $k=1$, while variability of the quarter-hourly scale corresponds to the harmonic $k=4$, i.e., four oscillation cycles per hour. Hence, the most relevant question is how $E_{h,1}$ and $E_{h,4}$ change after the market reform.

\section{Results}
\label{sec:results}

\subsection{Power spectral density of the frequency time series}
\label{sec:results-psd}

\begin{figure}
    \centering
    \includegraphics[width=\columnwidth]{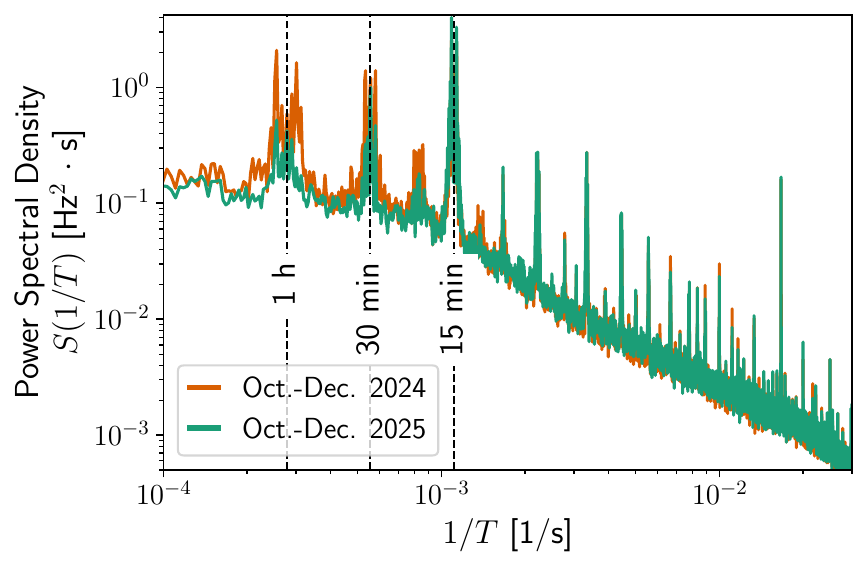}
    \caption{Power spectral density $S(1/T)$ of the grid frequency time series before (orange) and after (green) the market reform. We plot $S$ versus $1/T$, where $T$ denotes the period.
    The characteristic peaks corresponding to oscillations with hourly and half-hourly periods appear to be suppressed after the market change in 2025.}
    \label{fig:psd_frequency}
\end{figure}

To highlight the change in characteristic oscillations in the grid frequency, we calculate its power spectral density as described in Sec.~\ref{sec:method_psd}.
This provides a global frequency-domain characterization of variability across all time scales. 

Figure~\ref{fig:psd_frequency} shows the power spectral density of the grid frequency before and after the market reform. Prior to the reform, pronounced spectral peaks are observed at periods of $T = 1\,\si{h}$ and $T = 30\,\si{min}$, reflecting strong systematic hourly and half-hourly variability. After the reform, these peaks are substantially reduced, indicating a marked suppression of market-induced frequency fluctuations on these time scales. In contrast, the spectral peak at $T = 15\,\si{min}$ increases, consistent with the growing importance of quarter-hourly structures following the introduction of quarter-hourly trading intervals in the day-ahead market. Notably, further strong peaks emerge at lower periods due to control actions~\cite{mishra2025understanding,hartmann2025cyber}.

\subsection{Daily profile, histogram and long-term trends}
\label{sec:res_profile_hist}

\begin{figure}
    \centering
    \includegraphics[width=\columnwidth]{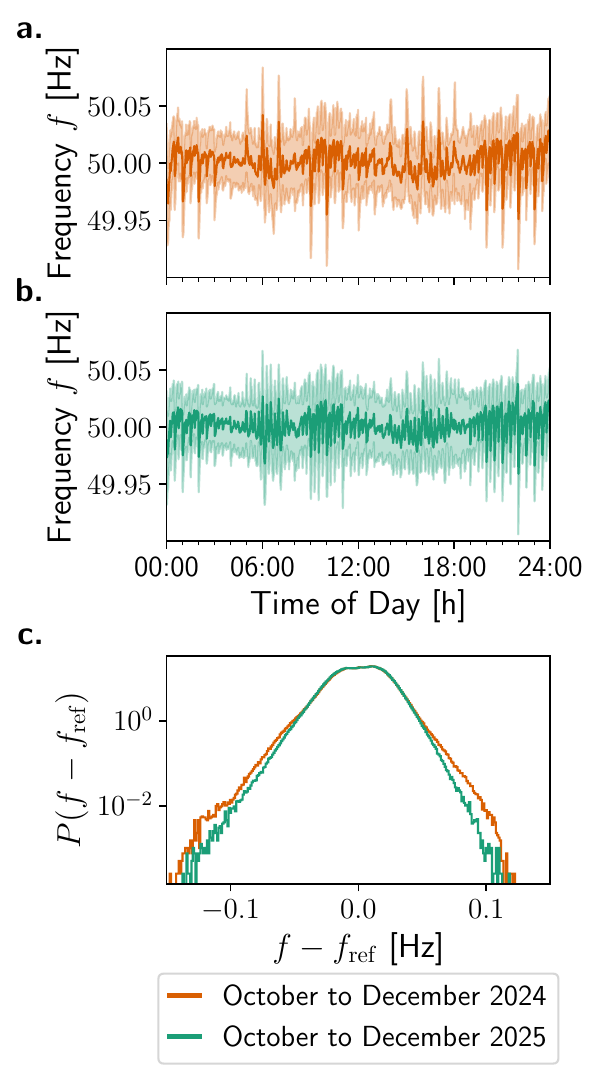}
    \caption{Analysis of the impact of the market reform on the global statistical properties of the grid frequency.
    (a,b) Daily profiles of the grid frequency $f(t)$ before (orange ) and after (green) the market reform on 1 October 2025 (see Sec.~\ref{sec:method-daily-profile} for details). Large deterministic frequency deviations are substantially reduced after the market reform.
    (c) Histogram of the frequency deviation $\tilde f(t) = f(t) - f_{\rm ref}$ before (orange) and after (green) the market reform. The likelihood of frequency deviations exceeding $5\,\si{mHz}$ is strongly reduced.
    }
\label{fig:daily_profiles}
\end{figure}

\begin{figure}    \centering
    \includegraphics[width=\columnwidth]{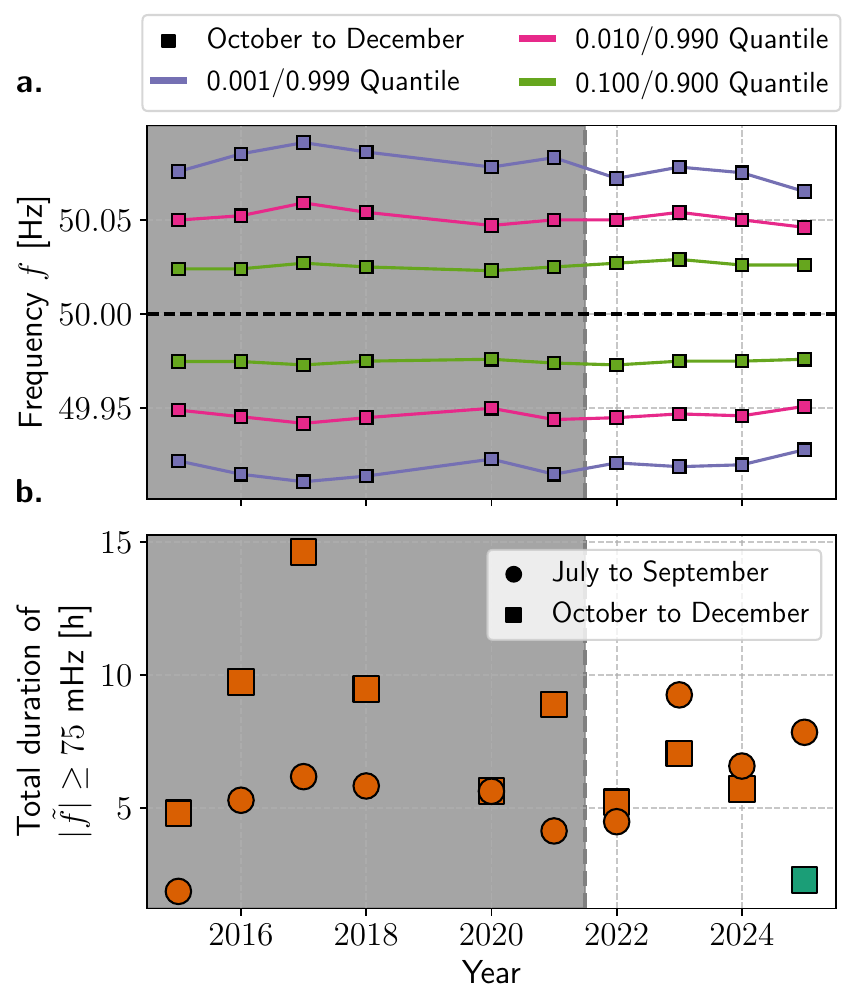}
    \caption{
    Long-term trends of frequency deviations.
    Upper panel: Evolution of selected quantiles of the frequency distribution during the months October--December from 2015 to 2025. 
    Lower panel: Total duration of periods with large frequency deviations $|\tilde f(t)| \ge 75\,\si{mHz}$. We provide data for the months October--December as well as July--September for comparison. The grey shading indicates that data publication changed in June 2022 (see. Sec.~\ref{sec:data_source}).
    }   
\label{fig:long-term}
\end{figure}

The daily frequency profiles shown in Fig.~\ref{fig:daily_profiles} (a,b) indicate that the market reform on 1 October 2025 has substantially reduced large deterministic frequency deviations in the Central European synchronous area. In particular, the pronounced positive spikes observed in the morning and early evening before the reform are strongly diminished. Negative deviations are also reduced significantly, although they remain clearly visible at certain times of the day. Moreover, hourly and quarter-hourly features appear to be of comparable magnitude in the post-reform period, consistent with the transition from hourly to quarter-hourly trading blocks in the day-ahead market.

A notable exception occurs at 22:00, where a pronounced negative frequency deviation persists even after the market reform. A plausible explanation for this feature is the shutdown or throttling of wind turbines to comply with nighttime noise regulations. In Germany, for example, stricter noise limits apply during the nighttime period between 22:00 and 6:00~\cite{talarm1998}. A comparison of such regulations across European countries is provided in Ref.~\cite{nieuwenhuizen2015differences}. This suggests that the remaining spike at 22:00 is at least partly driven by regulatory constraints on wind generation and is therefore largely unaffected by the market reform.

These observations are corroborated by the histogram of the frequency deviations shown in Fig.~\ref{fig:daily_profiles}(c). The likelihood of large frequency deviations is substantially reduced after the market reform, particularly for positive deviations. ENTSO-E defines $75\,\si{mHz}$ as the maximum acceptable magnitude of deterministic frequency deviations~\cite{entsoe_dfd2020}. Using this threshold, the cumulative duration of intervals with $\tilde f \geq 75\,\si{mHz}$ decreased from $2.23$ hours to $0.61$ hours ($-72.70\%$) over the considered three--month period. Likewise, the cumulative duration of intervals with $\tilde f \leq -75\,\si{mHz}$ decreased from $3.49$ hours to $1.70$ hours ($-51.27\%$).

\begin{figure*}[tb]
    \centering
    \includegraphics[width=\linewidth]{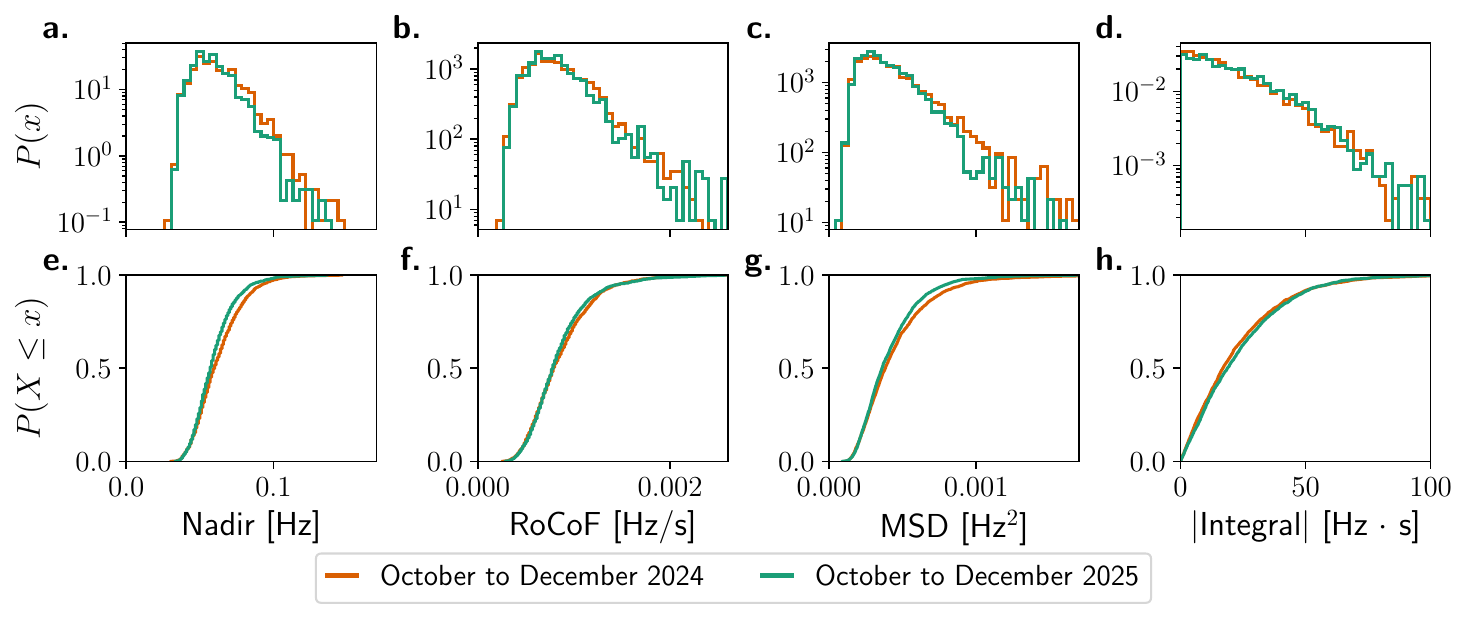}
    \caption{Distributions of stability indicators before and after the market reform on 1 October 2025.
    The probability density $P(x)$ of the different stability indicators and the cumulative distribution function can be found in the top row and bottom row, respectively.
    The definition of the indicators is summarized in Sec.~\ref{sec:methods-indicators}.  
    }
    \label{fig:stability_indicators_hourly}
\end{figure*}

To place these short-term observations into a longer-term perspective, Fig.~\ref{fig:long-term}(a) shows selected quantiles of the frequency deviation histogram for the period from 2015 to 2025, focusing on the months October to December in each year. The most extreme quantiles (1\% and 99\%), as well as the even rarer tails (0.1\% and 99.9\%), are shifted toward the mean in 2025 compared to 2024, consistent with the reduction of large frequency deviations observed above. Notably, these quantiles attain their smallest absolute values in 2025 among all years under consideration, indicating that extreme frequency deviations during the period of interest are less pronounced than in any previous year since 2015. This finding supports the conclusion that the market reform was effective in mitigating large deterministic frequency deviations. At the same time, we note that reductions of comparable magnitude were also observed in earlier years, for instance between 2017 and 2018, highlighting that year-to-year variability can be substantial.

As a complementary long-term indicator, we analyze the cumulative duration of periods with $|\tilde f(t)| \ge 75\,\si{mHz}$ in Fig.~\ref{fig:long-term}(b). During the period October to December 2025, the cumulative duration above this threshold amounts to $2.31$ hours, which is substantially lower than in the corresponding periods of all years since 2015. The largest value is observed in 2017, exceeding $14$ hours. For comparison, we also show the same metric for the months July to September. In contrast to the October–December period, the year 2025 exhibits a comparatively high cumulative duration of frequency deviations exceeding $75\,\si{mHz}$, amounting to slightly less than 8 hours for the July–September period. This suggests that the pronounced reduction observed in the period of interest is unlikely to be driven by longer-term trends or structural changes unrelated to the market reform.

In summary, the long-term analysis indicates that the market reform was effective in reducing the occurrence of deterministic frequency deviations exceeding the critical threshold of $75\,\si{mHz}$ during the months following its implementation.

\subsection{Indicators of frequency deviations}

We now analyze the indicators characterizing frequency deviations introduced in Sec.~\ref{sec:methods-indicators}. Results are shown in Fig.~\ref{fig:stability_indicators_hourly}.
The most pronounced changes are observed for the nadir. The likelihood of large nadir values is clearly reduced after the market reform. In particular, the total number of hours in which the absolute nadir exceeds $75\,\si{mHz}$ decreases from 458 to 288 (-37.12\%), while the number of hours with nadir values exceeding $100\,\si{mHz}$ decreases from 62 to 39 (-37.10\%). 
Consistent with the histogram of the raw frequency time series seen in Fig.~\ref{fig:daily_profiles}, the reduction of frequency deviations can be seen to affect nadir values above around $40\,\si{mHz}$, suggesting that the market reform was effective in mitigating frequency deviations of intermediate magnitudes, while its impact on small and rare events is more limited.

Changes in the RoCoF are also clearly visible but less pronounced. This is noteworthy, as the RoCoF is directly related to power imbalances through the aggregated swing equation~\cite{ulbig2014impact}. The most significant differences are observed for RoCoF values between approximately $0.8\,\si{mHz/s}$ and $1.5\,\si{mHz/s}$, corresponding to large but not extreme events. This indicates that the market reform was successful in mitigating ordinary large RoCoF excursions, whereas very large RoCoF events persist. A plausible explanation is that such extreme events are not primarily driven by typical market-induced imbalances but by other disturbances.

The mean square deviation (MSD), which reflects the total primary control effort and thus operational costs~\cite{tyloo2020primary}, exhibits a modest reduction of large values after the market reform. This points to a slight decrease in the average activation of primary control.

By contrast, no significant change is observed for the integrated frequency deviation. This is consistent with the fact that the integral primarily captures long-lasting power imbalances rather than rapid ramps. Such persistent imbalances can arise from forecasting errors or slow solar power ramps~\cite{kruse2021revealing}, which are largely independent of the market design.

\subsection{Principal component analysis}

\begin{figure}[bt]
    \centering
    \includegraphics[width=\columnwidth]{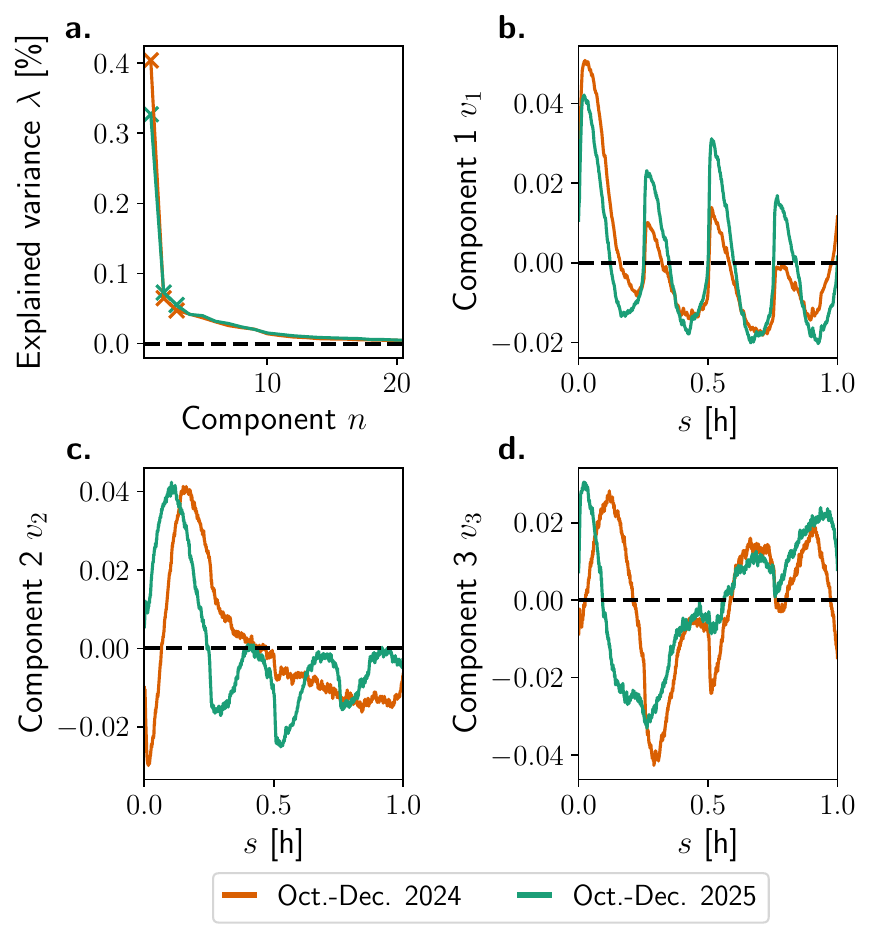}
    \caption{Principal Component analysis of the frequency time series $\tilde f(t)$ before (orange) and after (green) the market reform.
    (a) Explained variance $\lambda_n$ per principal component. The contribution of the first component has decreased from approx.~40.40 \% to 32.71 \%.
    (b-d) Shape of the three leading components $v_n(s)$, i.e., $n=1,2,3$, over the time $s$ within the hourly segments.   
    }
    \label{fig:pca-modes}
\end{figure}

The principal component analysis reveals pronounced changes in the structure of systematic intra-hour frequency variations following the market reform (Fig.~\ref{fig:pca-modes}). The shape of the leading component $v_1(s)$ illustrates the growing importance of quarter-hourly trading most clearly: the spikes occurring after 15, 30, and 45 minutes have increased substantially, while the spike at the beginning of the full hour has diminished. At the same time, the fraction of variance explained by this component has decreased, indicating that the dominant intra-hour pattern is less pronounced and that a richer substructure of frequency variations contributes to the overall dynamics. This observation is consistent with smaller marketdesign-induced generation ramps  at the beginning of the full hour.

Changes in the second and third principal components are more subtle but reveal an important restructuring of intra-hour frequency patterns following the market reform. Notably, the roles of the second and third components are partly exchanged. Prior to the reform, the second component primarily encoded temporal shifts of the dominant frequency spike at the beginning of the full hour. Specifically, a positive product of the corresponding coefficients $a_{h,1} a_{h,2} > 0$ shifted the peak toward later times, while $a_{h,1} a_{h,2} < 0$ resulted in an earlier occurrence. This component exhibited little substructure related to quarter-hourly trading, whereas the third component before the reform already displayed a clear quarter-hourly pattern.

After the market reform, the second principal component exhibits a pronounced quarter-hourly structure, with spikes at the beginning of the hour and at 15, 30, and 45 minutes having opposite signs. Consequently, this component captures variations in the relative importance of frequency deviations associated with hourly and quarter-hourly generation ramps. By contrast, the third component now primarily encodes temporal shifts of the hourly peak toward earlier or later times. It has a weak but clearly visible quarter-hourly substructure.

These findings are consistent with the growing importance of quarter-hourly trading in shaping systematic frequency variations. In particular, the second component is required to describe shifts between hourly- and quarter-hourly-dominated frequency patterns, which aligns with the observation that some hourly peaks are affected more strongly than others, with the 22:00 spike providing a prominent example.

\subsection{Fourier-based functional data analysis}
\label{sec:results-fourier}

For a more detailed, scale-resolved analysis, we turn to the Fourier-based functional data analysis described in Sec.~\ref{sec:methods-fourier}. Each hourly segment of the centered frequency time series $\tilde f_h(s)$ is decomposed into Fourier components according to Eq.~\ref{eq:Fourier-expand}. In contrast to the principal component analysis, which extracts data-driven modes, this approach employs a fixed set of sine and cosine basis functions, allowing individual temporal scales to be analyzed directly.

\begin{figure}
    \centering
    \includegraphics[width=\columnwidth]{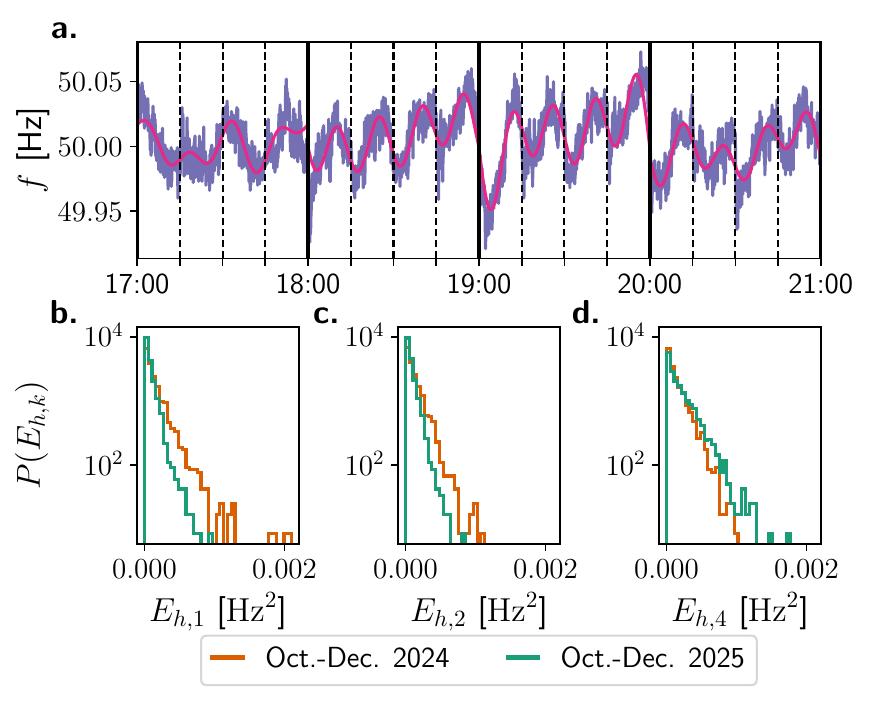}
    \caption{
    Fourier-based functional data analysis.
    (a) Expansion of hourly frequency time series into Fourier modes up to $k=4$, shown as a solid line, with the frequency time series behind it. 
    (b-d) Histogram of the energy $E_{h,k}$ in the modes $k=1,2,4$ capturing 
    variations on the time scale of hours, half-hours and quarter-hours, respectively.  
    }
    \label{fig:fourier-expansion}
\end{figure}

Figure~\ref{fig:fourier-expansion} illustrates the Fourier expansion. Retaining Fourier modes up to order $k=4$ captures slow intra-hour variations of the frequency and is therefore well suited to describe the dominant temporal structure associated with deterministic frequency deviations.

The lower panels of Fig.~\ref{fig:fourier-expansion} show the distributions of the oscillatory energy $E_{h,k}$ in the modes $k=1$, $k=2$, and $k=4$ across all hourly intervals. We observe that the energy associated with the hourly ($k=1$) and half-hourly ($k=2$) modes is substantially reduced after the market reform. Notably, the reduction is particularly pronounced for large values of $E_{h,1}$, indicating a strong suppression of large hourly-scale frequency variations.
This is highlighted by evaluating the 90-percent quantile $q_{90\%}$ for $E_{h,1}$ and $E_{h,2}$. 
While $q_{90\%}$ was around $351.51\,\si{mHz}^2$ and $308.77\,\si{mHz}^2$ before the market change, it decreased to $184.67\,\si{mHz}^2$ and $180.22\,\si{mHz}^2$ for the time interval after the market change for $E_{h,1}$ and $E_{h,2}$, respectively.  

At the same time, the energy of the quarter-hourly mode ($k=4$) increases, consistent with the shift from hourly to quarter-hourly trading intervals.
The 90-percent quantile increased from $358.85\,\si{mHz}^2$ to $466.95~\si{mHz}^2$, which shows that $E_{h,4}$ takes larger values for the time period after the market change. 
These findings are consistent with the previous results obtained from the daily profiles and the principal component analysis: characteristic hourly variability is strongly reduced, while quarter-hourly variability gains importance. We note, however, that these results cannot be directly linked to extreme nadir values, which are sensitive to faster temporal fluctuations and higher Fourier modes not considered in the present analysis.
Furthermore, these results on the change in distribution of oscillatory energy within individual hourly intervals together with the changing global spectral results on the full time series presented in Sec.~\ref{sec:results-psd}, provide complementary evidence for a shift from hourly- to quarter-hourly-dominated frequency variability after the market reform.

\section{Conclusion and Outlook}
\label{sec:discussion}

In this paper, we analyzed the impact of the European day-ahead market reform on 30 September 2025 on deterministic frequency deviations (DFDs) in the Central European power system. By comparing frequency measurements before and after the reform, we combined time-domain indicators, daily frequency profiles, principal component analysis, Fourier-based functional data analysis, and power spectral density analysis to assess changes in characteristic intra-hour frequency patterns. This multi-perspective approach allowed us to quantify both global and scale-specific changes in frequency deviations associated with the market reform.

Our results show that the market reform was largely successful in mitigating deterministic frequency deviations, in particular the characteristic deviations occurring at the beginning of each full hour. Hourly-scale variability is substantially reduced across all analyses, while quarter-hourly structures gain relative importance, consistent with the transition from hourly to quarter-hourly trading intervals. However, DFDs are not the sole contributors to large frequency deviations, and several effects remain largely unchanged. Notably, the pronounced frequency dip around 22:00 persists after the reform, plausibly linked to regulatory constraints such as nighttime noise limits for wind generation. In addition, long-lasting power imbalances related to forecasting errors are largely unaffected. As a result, while the frequency quality has improved overall, large deviations are not fully eliminated, which is reflected in the statistics of the hourly nadir.

These findings are consistent with, and further supported by, recent observations indicating that finer temporal resolution alone is not sufficient to fully eliminate market-induced ramping and balancing challenges. While our analysis focuses on frequency outcomes, complementary evidence from the day-ahead market provides insight into the underlying scheduling and bidding behavior. Following the transition of the European day-ahead electricity market to a 15-minute resolution, pronounced intra-hour price variability has emerged, characterized by recurring saw-tooth patterns and persistently high intra-hour price spreads that were previously mainly observed in intraday trading \cite{sdac15min2025}. These price signatures indicate that physical flexibility at the 15-minute level remains insufficient, such that the expected smoothing effects of finer scheduling are only partially realized. In particular, variable renewable generators have adapted their bids to the 15-minute market structure, whereas conventional thermal units rely more heavily on linked block offers spanning multiple 15-minute intervals to avoid short-duration start-up costs, thereby reintroducing temporal aggregation at the market level. Consistent with this interpretation, order book data show that a considerable share of block orders continues to start at the beginning of an hour and end at the end of an hour, effectively preserving hourly scheduling structures despite the finer market time resolution \cite{sdac15min2025}. As a result, interactions between bidding behavior, market liquidity, and physical constraints can sustain synchronized ramps that may still translate into residual DFDs.

Looking ahead, frequency control and frequency performance remain critical challenges during the ongoing transition toward renewable-dominated power systems with low rotational inertia~\cite{milano2018foundations}. Past events have shown that particularly critical situations can arise when DFDs coincide with contingencies. For example, in January 2019 a technical error in the load–frequency control system, combined with a pronounced DFD, led to an exceptionally large frequency deviation and ultimately to load shedding of approximately 1.5\,GW in the Continental European synchronous area~\cite{entsoe2019continental}. Even more severe risks emerge when DFDs coincide with major contingencies, underscoring the need for continued research on synthetic inertia, advanced converter control, and fast-acting frequency support from power-electronic devices. Beyond these technical measures, our results highlight that economic regulations and market design exert a strong influence on the technical operation of power systems and must therefore be considered an integral part of future frequency-control strategies.

\bibliographystyle{ieeetr}

\end{document}